\documentclass[nofootinbib,preprint,preprintnumbers]{revtex4}
\usepackage{epsfig,footnote}
\usepackage{ulem}
\usepackage{color}
\usepackage{array}
\usepackage{amssymb}
\usepackage{amsmath}
\usepackage{graphicx,subfigure}
\usepackage{longtable}
\usepackage{verbatim}
\usepackage{amsfonts}
\usepackage{hyperref}
\usepackage{cancel}
\usepackage{multirow}
\usepackage{textcomp}

\begin{document}

\title{$K_{S}^{0}-K_{L}^{0}$ asymmetries in $D$-meson decays}

\author{Di Wang$^{1}$
, Fu-Sheng Yu$^{1,2}$
, Peng-Fei Guo$^{1}$, and Hua-Yu Jiang$^{1}$}
\address{%
$^1$ School of Nuclear Science and Technology,  Lanzhou University,  Lanzhou 730000,  People's Republic of China \\
$^{2}$Research Center for Hadron and CSR Physics, Lanzhou University
and Institute of Modern Physics of CAS, Lanzhou 730000, People's Republic of China
}

\begin{abstract}
The $K_{S}^{0}-K_{L}^{0}$ asymmetries in the $D$ meson decays, induced by the interference between the Cabibbo-favored and the doubly Cabibbo-suppressed amplitudes,  can help to understand the dynamics of charm decays.
All possible processes of two-body non-leptonic $D$ decays into one neutral kaon and another pseudoscalar or vector meson are considered.
We study the $K_{S}^{0}-K_{L}^{0}$ asymmetries and the branching fractions of corresponding processes in the factorization-assisted topological-amplitude approach in which significant flavor $SU(3)$ symmetry breaking effects are included.
The branching fractions of $K_{L}^{0}$ modes are predicted.
It is first found that the $K_{S}^{0}-K_{L}^{0}$ asymmetries in the $D^0$-meson decays are shifted by the $D^0-\overline D^0$ mixing parameter $y_{D}\simeq0.006$, to be $0.113\pm0.001$ for all the relevant $D^{0}$ decay modes.
Our results on $K_{S}^{0}-K_{L}^{0}$ asymmetries are consistent with the current data and could be tested by experiments in the future.
\end{abstract}

\maketitle

\section{Introduction}
The study of $D$-meson decays and mixing can provide some useful information with respect to flavor mixing and $CP$ asymmetries~\cite{lab01}.
The two-body nonleptonic decays of $D$ mesons can be classified into three types: Cabibbo-favored (CF), singly Cabibbo-suppressed (SCS), and doubly Cabibbo-suppressed (DCS) processes. In the Standard Model (SM), they are, respectively, corresponding to the Cabibbo-Kobayashi-Maskawa (CKM) matrix elements, $|V_{cs}^{*}V_{ud}|\sim1$, $|V_{cd}^{*}V_{ud}|\sim |V_{cs}^{*}V_{us}|\sim\lambda$, and $|V_{cd}^{*}V_{us}|\sim\lambda^{2}$ with the Wolfenstein parameter $\lambda=\sin\theta_{C}\approx0.225$ and $\theta_{C}$ as the Cabibbo angle. Unlike the CF and SCS processes mostly observed in experiments, only a few DCS modes are well measured due to the relatively small branching fractions~\cite{PDG}. However, the studies on DCS processes have great interests for us. Because of the  relative smallness in the SM, the DCS processes can be significantly affected by new physics beyond the Standard Model. For example, there would be new $CP$ violating effects in the DCS processes in some new physics models, thereby affecting the determination of mixing parameters and indirect $CP$ phases in the $D^{0}-\overline D^{0}$ system  \cite{HFAG,Kagan:2009gb}. Besides, the studies on DCS processes can help to test the flavor $SU(3)$ symmetry and understand the dynamics of charmed hadron decays and the mechanism of $D^{0}-\overline D^{0}$ mixing~\cite{FAT1,FAT2,Cheng:2010ry,Cheng:2016ejf,Cheng:2012xb,Muller:2015lua,Bhattacharya:2008ss,Bhattacharya:2008ke,Bhattacharya:2009ps,Donoghue:1985hh,Buccella:1996uy,lab3,Falk:2001hx,Cheng:2010rv}.

Among the DCS modes,  the decaying of $D$ mesons into $K^0$ in the final states is actually involved in the processes with $K_{S}^{0}$, which are dominated, however, by CF modes of $D$ decaying into $\overline K^{0}$.  We cannot distinguish the effects of the CF and DCS amplitudes in the individual data of $D\to K_{S}^{0}f$. In some of the literatures, the decays with $K_{S}^{0}$ in the final states are always approximately considered as saturated by CF contributions, and hence the DCS information is neglected in such processes~\cite{Cheng:2010ry,Cheng:2016ejf,Cheng:2012xb,Rosner:1999xd}.

 The difference between the $K^0_S$ and $K^0_L$ modes induced by interference between the CF and DCS amplitudes was first pointed out by Bigi and Yamamoto~\cite{Bigi}. They proposed the observable of the $K_{S}^{0}-K_{L}^{0}$ asymmetry to describe the difference between modes with $K_{S}^{0}$ and $K_{L}^{0}$. In the two-body decays of $D\to K^0_{S,L}f$ with $f$ as the other meson in the final state except for the neutral kaons, the $K_{S}^{0}-K_{L}^{0}$ asymmetries are defined by
\begin{equation}\label{Rf}
   R(f)\equiv\frac{\Gamma(D\rightarrow K_S^0f) -\Gamma(D\rightarrow K_L^0f)}{\Gamma(D\rightarrow K_S^0f) + \Gamma(D\rightarrow K_L^0f)}.
 \end{equation}
The nonvanishing values of $R(f)$ would be induced by the interference between the decay amplitudes of $D\to \overline K^{0}f$ (CF transitions) and $D\to K^{0}f$ (DCS transitions). Therefore, the determination on the $K_{S}^{0}-K_{L}^{0}$ asymmetries in $D$ decays can be useful to study the DCS processes.
The asymmetries in $D\to K_{S,L}^{0}\pi$ decays have been measured by the CLEO Collaboration~\cite{lab10}\footnote{The BESIII collaboration has reported their preliminary result with only statistical error that~\cite{RBES}
\begin{equation}
  R(D^0\to K_{S,L}^{0}\pi^0)=0.1077\pm0.0125. \nonumber
\end{equation}}
\begin{equation}
\begin{split}
  R(D^0\to K_{S,L}^{0}\pi^0)=0.108\pm0.025\pm0.024,\\
  R(D^+\to K_{S,L}^{0}\pi^+)=0.022\pm0.016\pm0.018.
  \end{split}
\end{equation}

The $K_{S}^{0}-K_{L}^{0}$ asymmetries have been studied in the QCD factorization approach~\cite{Gao:2006nb,lab4}. However, since the charm quark mass is not heavy enough, the QCD-inspired methods, such as the QCD factorization approach~\cite{lab5}, the perturbative QCD approach~\cite{lab6}, and the soft-collinear effective theory~\cite{lab02}, are not suitable for charmed hadron decays.
The asymmetries are also predicted in the conventional topological diagrammatic approach under the $SU(3)$ flavor symmetry~\cite{Bhattacharya:2008ss,Bhattacharya:2009ps,Cheng:2010ry}, but it is known that the $SU(3)$ breaking effects can be as large as $30\%$ in charm decays and, thus have to be considered.
 In~\cite{Muller:2015lua}, the authors studied the $K_{S}^{0}-K_{L}^{0}$ asymmetries in the topological approach including linear $SU(3)$ breaking effects.  Since there are too many parameters in this method, the predictive power is limited.

In this work, we study the $K_{S}^{0}-K_{L}^{0}$ asymmetries in the factorization-assisted topological-amplitude (FAT) approach~\cite{FAT1,FAT2}, in which nonperturbative contributions are included and significant flavor $SU(3)$ symmetry breaking effects are well expressed. It has been shown that the FAT approach works well in $D$ meson decays. The results on branching fractions are consistent with experimental data in the $D$ decays into two pseudoscalar mesons ($PP$), or one pseudoscalar meson and one vector meson $(PV)$. Furthermore, the prediction on the $CP$ asymmetry difference $\Delta A_{CP}=A_{CP}(D^{0}\to K^{+}K^{-})-A_{CP}(D^{0}\to \pi^{+}\pi^{-})$ in the FAT approach~\cite{FAT1} is verified by recent LHCb collaboration~\cite{Aaij:2016cfh}. The FAT approach will be introduced in details in the following sections.

In this paper, we will study the $K_S^0-K_L^0$ asymmetries in the SM in the $D\to PP$ decay modes of
\begin{equation*}
  D^0\rightarrow K_{S,L}^0\pi^0,    \quad  D^0\rightarrow K_{S,L}^0\eta, \quad  D^0\rightarrow K_{S,L}^0\eta^{\prime},  \quad
 D^+\rightarrow K_{S,L}^0\pi^+,    \quad
  D^+_s\rightarrow K_{S,L}^0K^+,
 \end{equation*}
 and firstly in the $D\rightarrow PV$ decay modes of
 \begin{equation*}
  \begin{split}
  D^0\rightarrow K_{S,L}^0\rho^0,    \quad  D^0\rightarrow K_{S,L}^0\omega, \quad D^0\rightarrow K_{S,L}^0\phi, \quad
 D^+\rightarrow K_{S,L}^0\rho^+,    \quad
  D^+_s\rightarrow K_{S,L}^0K^{*+}.
  \end{split}
 \end{equation*}
Furthermore, the $D^0-\overline{D}^0$ mixing effects will first be considered in $K_{S}^{0}-K_{L}^{0}$ asymmetries in the neutral $D$ meson decay modes, which we find to be non-negligible.

The plan of this paper is as follows.
In Sec.~\ref{formalism}, we will show the general formulas of the $K_{S}^{0}-K_{L}^{0}$ asymmetries, $R(f)$, and the $D^{0}-\overline D^{0}$ mixing effects on it. In Sec.~\ref{FAT}, we shall present the amplitude decompositions of $D\to PP$ and $D\to PV$ decays in the FAT approach.  The numerical results on branching fractions and $K_{S}^0-K^0_L$ asymmetries will be given in Sec.~\ref{result}. Sec.~\ref{CON} is the conclusion.

\section{The Formalism of the $K_{S}^{0}-K_{L}^{0}$ asymmetries }\label{formalism}

\subsection{The $K_S^0-K_L^0$ asymmetries in charged $D$ decays}

The $K_S^0$ and $K_L^0$ states are linear combinations of $K^{0}$ and $\overline K^{0}$, under the convention of $\mathcal{CP}|K^0\rangle = -|\overline{K}^0\rangle$, as
 \begin{equation}\label{eq:KSKL}
  \begin{split}
|K_{S}^0\rangle  &=   \frac{1}{\sqrt{2(1+|\epsilon|^2)}}\left[(1+\epsilon)|K^0\rangle-(1-\epsilon)|\overline{K}^0\rangle\right], \\
|K_{L}^0\rangle  &=   \frac{1}{\sqrt{2(1+|\epsilon|^2)}}\left[(1+\epsilon)|K^0\rangle+(1-\epsilon)|\overline{K}^0\rangle\right],
\end{split}
 \end{equation}
where $\epsilon=|\epsilon|e^{i\phi_\epsilon}$ is a small complex parameter indicating the indirect $CP$ violating effect, with the value of $|\epsilon|=(2.228\pm0.011)\times10^{-3}$ and $\phi_\epsilon=43.5^\circ\pm0.5^\circ$~\cite{PDG}. We start with this more general formula of $K^{0}-\overline K^{0}$ mixing, and will find later that the parameter $\epsilon$ is negligible in the $K_S^0-K_L^0$ asymmetries.

In order to study the $K_S^0-K_L^0$ asymmetries in Eq.(\ref{Rf}), we express the amplitudes of $D\to \overline K^{0}f$ and $D\to K^{0}f$ decays as
\begin{align}\label{eq:ampCFDCS}
\mathcal{A}(D\to \overline K^{0}f)=\mathcal{T}_{\rm CF} e^{i(\phi_{\rm CF}+\delta_{\rm CF})},~~~~~\mathcal{A}(D\to  K^{0}f)=\mathcal{T}_{\rm DCS}\, e^{i(\phi_{\rm DCS}+\delta_{\rm DCS})},
\end{align}
where $\mathcal{T}_{\rm CF,\,DCS}$ are real, $\phi_{\rm CF,\,DCS}$ and $\delta_{\rm CF,\,DCS}$ are the weak and  strong phases for the CF and DCS amplitudes, respectively. The CKM matrix elements have been involved in $\mathcal{T}_{\rm CF,\,DCS}$.
From Eq.\eqref{eq:KSKL}, the amplitudes of the $D\rightarrow K_S^0f$ and $D\rightarrow K_L^0f$ decays are~\cite{x1}
\begin{equation}\label{eq:ampKSKL}
\begin{split}
 \mathcal{A}(D\rightarrow K_S^0f)  &   = \frac{1}{\sqrt{2(1+|\epsilon|^2)}}\left[(1+\epsilon^*)\,\mathcal{T}_{\rm DCS}\,e^{i(\phi_{\rm DCS}+\delta_{\rm DCS})} - (1-\epsilon^*)\,\mathcal{T}_{\rm CF}\,e^{i(\phi_{\rm CF}+\delta_{\rm CF})}\right], \\
\mathcal{A}(D\rightarrow K_L^0f)  &   = \frac{1}{\sqrt{2(1+|\epsilon|^2)}}\left[(1+\epsilon^*)\,\mathcal{T}_{\rm DCS}\, e^{i(\phi_{\rm DCS}+\delta_{\rm DCS})} +(1-\epsilon^*)\,\mathcal{T}_{\rm CF}\,e^{i(\phi_{\rm CF}+\delta_{\rm CF})}\right].
\end{split}
\end{equation}

For convenience, we define the ratio between the DCS and CF amplitudes as
 \begin{equation}
 \frac{\mathcal{A}(D\rightarrow K^0f)}{\mathcal{A}(D\rightarrow \overline{K}^0f)} = r_{f}\,e^{i(\phi_f+\delta_f)},
 \end{equation}
where $r_{f} = \mathcal{T}_{\rm DCS}/\mathcal{T}_{\rm CF}$, $\phi_f=\phi_{\rm DCS}-\phi_{\rm CF}$ and $\delta_f = \delta_{\rm DCS} - \delta_{\rm CF}$.
Note that $r_{f}$ is small, $r_{f}\propto|V_{cd}^{*}V_{us}/V_{cs}^{*}V_{ud}|\approx\lambda^{2}=\mathcal{O}(10^{-2})$. The parameters $r_{f}$ and $\delta_f$ depend on the individual processes and $\phi_f$ is mode independent in the SM. Then the $K_{S}^{0}-K_{L}^{0}$ asymmetries can be written as
\begin{align}\label{Rfepsilon}
R(f) &={|(1-\epsilon^{*})-(1+\epsilon^{*})\,r_{f}\,e^{i(\phi_f+\delta_f)}|^{2}-|(1-\epsilon^{*})+(1+\epsilon^{*})\,r_{f}\,e^{i(\phi_f+\delta_f)}|^{2}\over
|(1-\epsilon^{*})-(1+\epsilon^{*})\,r_{f}\,e^{i(\phi_f+\delta_f)}|^{2}+|(1-\epsilon^{*})+(1+\epsilon^{*})\,r_{f}\,e^{i(\phi_f+\delta_f)}|^{2}}
\nonumber\\
 &=-2\,r_{f}\frac{\cos(\phi_f+\delta_f)(1-|\epsilon|^2)+2\sin(\phi_f+\delta_f)Im(\epsilon)}{|1-\epsilon^{*}|^2+|1+\epsilon^{*}|^2\,r_{f}^2 }\nonumber\\
 &\simeq -2r_{f}\cos(\phi_f+\delta_f)-4r_{f}[Re(\epsilon)\cos(\phi_f+\delta_{f})+Im(\epsilon)\sin(\phi_f+\delta_{f})].
 \end{align}
The second term in the last line is sub-leading, at the order of $10^{-4}$, and hence can be safely neglected compared to the first term which is $\mathcal{O}(10^{-2})$.
Thus the $K_{S}^{0}-K_{L}^{0}$ asymmetries are not sensitive to the $CP$ violating effect in the $K^{0}-\overline K^{0}$ mixing system.
Besides, the weak phase difference $\phi_f$ is tiny in the SM,
$\phi_f=Arg\left[{V_{cd}^{*}V_{us}/V_{cs}^{*}V_{ud}}\right]$, $\sin\phi_f=\mathcal{O}(\lambda^{4})=\mathcal{O}(10^{-3})$.
Hence as a good approximation, the $K_{S}^{0}-K_{L}^{0}$ asymmetries can be expressed as
\begin{align}\label{Rfrf}
R(f) = -2\,r_{f}\cos\delta_f.
\end{align}
It can be expected that $R(f)=\mathcal{O}(10^{-2})$.
Therefore, the determination of the $K_{S}^{0}-K_{L}^{0}$ asymmetries is useful to understand the dynamics of the DCS decays, especially the relative strong phases between DCS and CF amplitudes.

Note that all the above formulas are in general for $D$ decays. In $D^{0}$ decay modes, the $D^0-\overline D^0$ mixing effects have to be considered, which will be discussed in the next subsection.
Besides, as seen above, the $CP$ violating effect in $K^{0}-\overline K^{0}$ mixing is negligible in the discussion of $K_{S}^{0}-K_{L}^{0}$ asymmetries. Thereby $K_{S}^{0}$ and $K_{L}^{0}$ are the $CP$ even and $CP$ odd states, respectively,
\begin{equation}\label{eq:KSKLCP}
|K_{S}^0\rangle  =   \frac{1}{\sqrt{2}}\left(|K^0\rangle-|\overline{K}^0\rangle\right), ~~~~~
|K_{L}^0\rangle  =   \frac{1}{\sqrt{2}}\left(|K^0\rangle+|\overline{K}^0\rangle\right).
 \end{equation}
In the following discussions, we will use the above formulas for $K_{S}^{0}$ and $K_{L}^{0}$ states, and the decay amplitudes of
\begin{equation}\label{ss1}
\begin{split}
 \mathcal{A}(D\rightarrow K_S^0f)  &
 =-\frac{1}{\sqrt{2}}\mathcal{T}_{\rm CF}\,e^{i\delta_{\rm CF}}\left(1-r_{f}e^{i\delta_{f}}\right) , \\
\mathcal{A}(D\rightarrow K_L^0f)  &
=\frac{1}{\sqrt{2}}\mathcal{T}_{\rm CF}\,e^{i\delta_{\rm CF}}\left(1+r_{f}e^{i\delta_{f}}\right).
\end{split}
\end{equation}
In the SM, there is a minus sign in $r_{f}=-\tan^{2}\theta_{c} \hat r_{f}$,
with $\hat r_{f}=|(\mathcal{T}_{\rm DCS}/\mathcal{T}_{\rm CF})(V_{cs}^{*}V_{ud}/V_{cd}^{*}V_{us})|$. Then the CF and DCS amplitudes would contribute  constructively (destructively) for the $K_{S}^{0}~(K_{L}^{0})$ modes in the case of $\cos\delta_{f}>0$, and conversely for $\cos\delta_{f}<0$. $R(f)$ can also be expressed as
\begin{align}\label{RfCabibbo}
R(f)=2\tan^{2}\theta_{C} ~\hat r_{f}\cos\delta_{f}.
\end{align}
Physics does not depend on the phase conventions. If $\mathcal{CP}|K^{0}\rangle=+|\overline K^{0}\rangle$, Eq.~\eqref{Rfrf} would become $R(f)=2r_{f}\cos\delta_{f}$. But as shown in~\cite{Cheng:2010ry} that the decay constants of  $K^{0}$ and $\overline K^{0}$ are opposite in sign in this case,
 there would be additional opposite sign between $\mathcal{A}(D\to K^{0}f)$ and $\mathcal{A}(D\to \overline K^{0}f)$, so then $r_{f}=\tan^{2}\theta_{c} \hat r_{f}$. Thus (\ref{RfCabibbo}) still holds under different phase conventions.

\subsection{The effect of the $D^0-\overline{D}^0$ mixing}

We will study the $D^0-\overline{D}^0$ mixing effect in the $D^0\to K_{S,L}^0f_{CP}^0$ decays, where $f_{CP}^0$ is a $CP$ eigenstate such as $\pi^{0}$, $\eta^{(\prime)}$, $\rho^{0}$, $\omega$ and $\phi$.
Under the convention of $\mathcal{CP}|D^{0}\rangle=-|\overline D^{0}\rangle$, the mass eigenstates of the neutral $D$ mesons can be written as
$|D^0_{1,2}\rangle = p|D^0\rangle \mp q|\overline{D}^0\rangle$ with $q/p=|q/p|e^{i\phi_D}$.
Some standard notations are used in neutral $D$-meson mixing:
\begin{equation}
  \begin{split}
   \mathcal{A}_{K_S^0}    & \equiv \mathcal{A}(D^0\to K_S^0f_{CP}^0), \quad \overline{\mathcal{A}}_{K_S^0}  \equiv \mathcal{A}(\overline{D}^0\to K_S^0f_{CP}^0),\\
    \mathcal{A}_{K_L^0}   &  \equiv \mathcal{A}(D^0\to K_L^0f_{CP}^0), \quad \overline{\mathcal{A}}_{K_L^0}  \equiv \mathcal{A}(\overline{D}^0\to K_L^0f_{CP}^0),\\
 \lambda_{K_S^0} &  \equiv \frac{q}{p}\frac{\overline{\mathcal{A}}_{K_S^0}}{\mathcal{A}_{K_S^0}}, \quad  \lambda_{K_L^0} \equiv \frac{q}{p}\frac{\overline{\mathcal{A}}_{K_L^0}}{\mathcal{A}_{K_L^0}}, \quad
  \Gamma_{D^0}   \equiv \frac{\Gamma_{D^0_1} + \Gamma_{D^0_2}}{2},  \\
   x_{D}  \equiv & \frac{\Delta m_{D^0}}{\Gamma_{D^0}}  = \frac{m_{D^0_1} - m_{D^0_2}}{\Gamma_{D^0}}, \quad y_{D} \equiv \frac{\Delta \Gamma_{D^0}}{2 \Gamma_{D^0}} = \frac{\Gamma_{D^0_1} - \Gamma_{D^0_2}}{2\Gamma_{D^0}},
  \end{split}
\end{equation}
where the amplitudes of $\overline{D}^0\rightarrow K_S^0f^0_{CP}$ and $\overline{D}^0\rightarrow K_L^0f^0_{CP}$ decays are expressed as
\begin{equation}\label{eq:ampbar}
\begin{split}
 \mathcal{A}(\overline{D}^0\rightarrow K_S^0f^0_{CP})  &   = -\eta_{CP}\,\eta_{K_{S}^{0}}\left[\mathcal{T}_{\rm DCS} e^{i(-\phi_{\rm DCS}+\delta_{\rm DCS})} - \mathcal{T}_{\rm CF}e^{i(-\phi_{\rm CF}+\delta_{\rm CF})}\right]/\sqrt2, \\
\mathcal{A}(\overline{D}^0\rightarrow K_L^0f^0_{CP})  &   =-\eta_{CP}\,\eta_{K_{L}^{0}}\left[ \mathcal{T}_{\rm DCS} e^{i(-\phi_{\rm DCS}+\delta_{\rm DCS})} + \mathcal{T}_{\rm CF} e^{i(-\phi_{\rm CF}+\delta_{\rm CF})}\right]/\sqrt2,
\end{split}
\end{equation}
  where the minus sign is from $\mathcal{CP}|D^0\rangle=-|\overline D^0\rangle$ and $\eta_{K_{S}^{0}}(\eta_{K_{L}^{0}})=+(-)$ from (\ref{eq:KSKLCP}), and $\eta_{CP}=\mathcal{PC}(-1)^{\mathcal{J}}$ with the quantum numbers $\mathcal{J}^{\mathcal{PC}}$ of $f_{CP}^{0}$. For the decay modes with pseudoscalar mesons of $\pi^{0}$, $\eta^{(')}$ or vector mesons of $\rho^{0}$, $\omega$ or $\phi$, the values of $\eta_{CP}=-1$.
In the absence of $CP$ asymmetry, we get the decay amplitudes of
\begin{equation}\label{ss2}
\begin{split}
 \mathcal{A}(\overline{D}^0\rightarrow K_S^0f^0_{CP}) &
 =-\frac{1}{\sqrt{2}}\mathcal{T}_{\rm CF}\,e^{i\delta_{\rm CF}}\left(1-r_{f}e^{i\delta_{f}}\right) , \\
 \mathcal{A}(\overline{D}^0\rightarrow K_L^0f^0_{CP}) &
=-\frac{1}{\sqrt{2}}\mathcal{T}_{\rm CF}\,e^{i\delta_{\rm CF}}\left(1+r_{f}e^{i\delta_{f}}\right).
\end{split}
\end{equation}
In the neutral $D$ meson system, $CP$ is conserved at the level of $10^{-4}$ so far~\cite{HFAG}. It is a good approximation that $|q/p|=1$ and $\phi_D=0$. So then, with Eqs~\eqref{ss1} and \eqref{ss2}, we get
\begin{equation}\label{lambda}
\lambda_{K_{S}^0} =1\qquad\text{and} \qquad \lambda_{K_{L}^{0}}=-1.
\end{equation}

The time-integrated decay rates of $D^0\to K_{S,L}^0f_{CP}^0$ decays can be expressed as~\cite{Grossman:2006jg}
\begin{equation}\label{eq:GammaD0}
  \begin{split}
  \Gamma(D^0\to K_{S,L}^0f_{CP}^0)
  =& \int_0^\infty \Gamma(D^0(t)\to K_{S,L}^0f_{CP}^0)dt
  \\
   = & \left|\mathcal{A}_{K_{S,L}^0}\right|^2 \bigg[1+{1+|\lambda_{K_{S,L}^0}|^2\over2}\frac{y_D^2}{1-y_D^2} - {1-|\lambda_{K_{S,L}^0}|^2\over2}\frac{x_D^2}{1+x_D^2}
   \\
   &  \quad\quad\quad\quad\quad  + Re(\lambda_{K_{S,L}^0})\frac{y_D}{1-y_D^2} - Im(\lambda_{K_{S,L}^0})\frac{x_D}{1+x_D^2}\bigg].
  \end{split}
\end{equation}
Notice that the first term  is independent from neutral $D$-meson mixing.
The $K_S^0-K_L^0$ asymmetries in the $D^0$ decays are
\begin{align}\label{eq:RD0}
R(f_{CP}^0) = -2r_{f}\cos\delta_{f} + y_{D}.
\end{align}
Compared to \eqref{Rfrf}, the mixing parameter $y_{D}$ contributes to $R(f_{CP}^0)$. The $y_{D}$ term in \eqref{eq:RD0} is attributed by the terms of $Re(\lambda_{K_{S,L}^0})$ in the rates $\Gamma(D^0\to K_{{S,L}}^0f_{CP}^0)$ in \eqref{eq:GammaD0}.
The current world averaging result of $y_{D}$ is $(0.62\pm0.08)\%$ in the case of $CP$ conservation~\cite{HFAG}.
If the precision of the measurements of $R(f^0_{CP})$ could reach up to $10^{-3}$, the neutral $D$-meson mixing effects have to be considered in the analysis of $D^0$ decays.

Similarly to the case of $K^{0}$ and $\overline K^{0}$, the phase conventions of $D^{0}$ and $\overline D^{0}$ would not affect the physical observables. If $\mathcal{CP}|D^{0}\rangle = + |\overline D^{0}\rangle$, an additional minus sign would exist in each equation of \eqref{eq:ampbar} and \eqref{lambda}, and the terms of $Re(\lambda_{K_{S,L}^{0}})$ and $Im(\lambda_{K_{S,L}^{0}})$ in \eqref{eq:GammaD0}. So then \eqref{eq:RD0} still holds in this convention.

\section{Amplitude decompositions in the FAT approach}\label{FAT}

The factorization-assisted topological-amplitude (FAT) approach works well for the charm decays~\cite{FAT1,FAT2}. It is based on the topological amplitudes according to the weak currents. There are four types of topological diagrams for the two-body nonleptonic $D$ meson decays at the tree level~\cite{qaz}: the color-favored tree emission amplitude $T$, the color-suppressed tree emission amplitude $C$, the $W$-exchange amplitude $E$, and the $W$-annihilation amplitude $A$, as shown in Fig.~$1$. Then the hypothesis of factorization is used, to calculate each topological
amplitude which is factorized into two parts: the short-distance Wilson coefficients and the long-distance hadronic matrix elements. The large nonperturbative and nonfactorizable contributions are parametrized to be determined by the experimental data. In this way, most $SU(3)$ flavor symmetry breaking effects are included.
\begin{figure}[phbt]\label{t1212}
\centering
\includegraphics[width=0.55\textwidth]{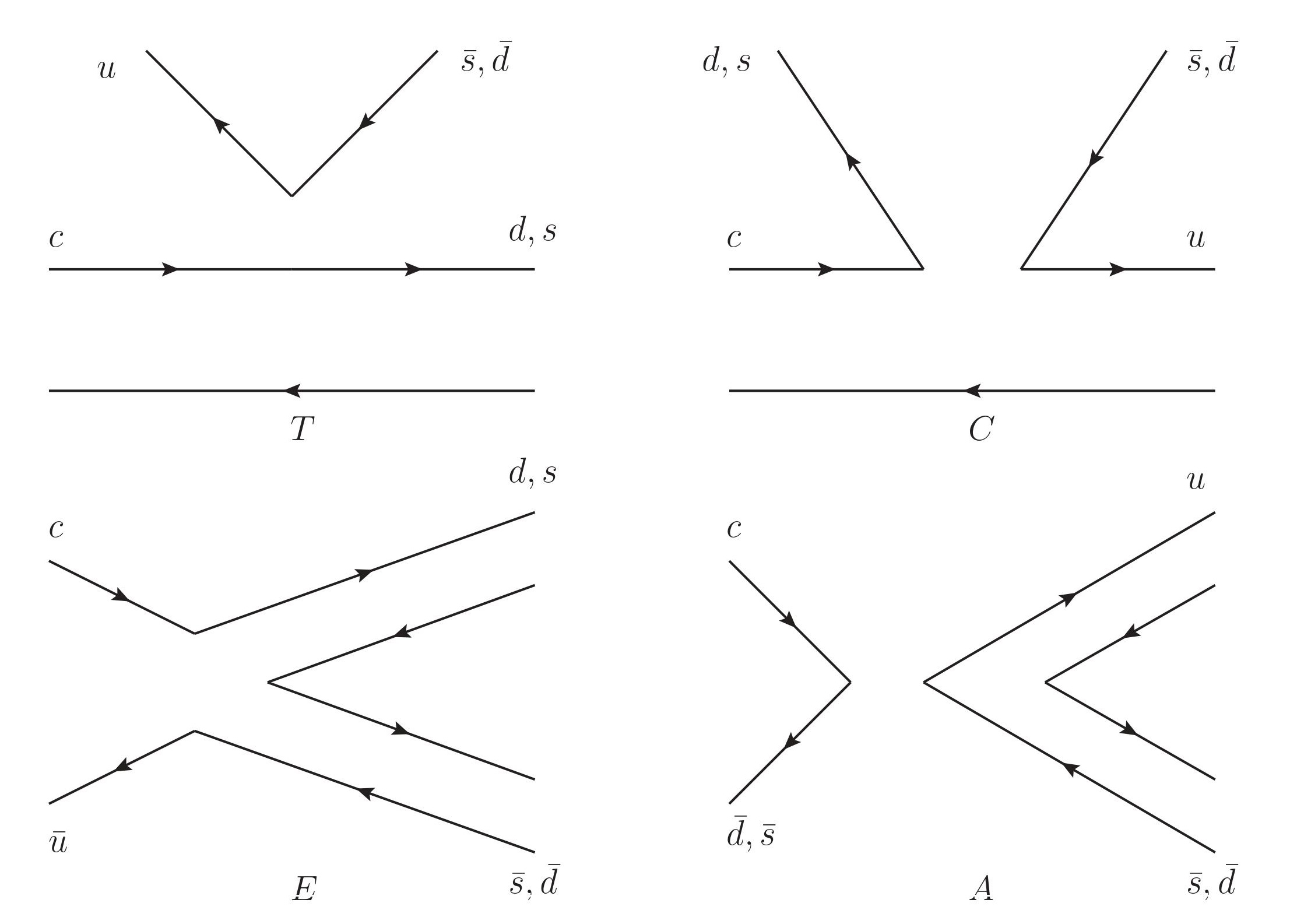}
\caption{
Four types of the topological diagrams contributing to two-body nonleptonic $D$ meson decays in the Standard Model: the color-favored tree amplitude $T$, the color-suppressed tree amplitude $C$, the $W$-exchange amplitude $E$ and the $W$-annihilation amplitude $A$.}
\end{figure}

The effective Hamiltonian of the charm decays in the SM can be written as~\cite{lab22}
\begin{equation}\label{Heff12}
\mathcal{H}_{eff}=\frac{G_F}{\sqrt{2}}
 V_{CKM}\left[C_1(\mu)Q_1(\mu)+C_2(\mu)Q_2(\mu)\right]+H.c.,
\end{equation}
where $G_F$ denotes the Fermi coupling constant, $V_{CKM}$ is the products of the Cabibbo-Kobayashi-Maskawa (CKM) matrix elements, $C_{1,2}$ are the Wilson coefficients. The current-current operators are written as
\begin{eqnarray}\label{pm1}
& Q_1=\bar{u}_{\alpha}\gamma_\mu(1-\gamma_5)q_{2\beta}
\bar{q}_{1\beta}\gamma^\mu(1-\gamma_5)c_{\alpha},~~~~~
Q_2=\bar{u}_{\alpha}\gamma_\mu(1-\gamma_5)q_{2\alpha}
\bar{q}_{1\beta}\gamma^\mu(1-\gamma_5)c_{\beta},
\end{eqnarray}
with $\alpha,\beta$ being the color indices, $q_{1,2}$ being the $d$ or $s$ quarks.

In the factorization hypothesis, the topological amplitudes in the $D\rightarrow PP$ modes can be written as~\cite{FAT1}
\begin{align}
T[C] &=  \frac{G_f}{\sqrt{2}}V_{CKM}a_{1}(\mu)[a_{2}(\mu)]f_{P_2}(m^2_D - m^2_{P_1})F_0^{D\rightarrow P_1}(m^2_{P_2}),\label{eq:TPP}
\\
E &=  \frac{G_f}{\sqrt{2}}V_{CKM}C_2(\mu)\chi^E_{q,s}e^{i\phi^E_{q,s}}f_D m^2_D \Big(\frac{f_{P_1}f_{P_2}}{f_\pi^2}\Big),\label{eq:EPP}
\\
A &=  \frac{G_f}{\sqrt{2}}V_{CKM}C_1(\mu)\chi^A_{q,s}e^{i\phi^A_{q,s}}f_D m^2_D \Big(\frac{f_{P_1}f_{P_2}}{f_\pi^2}\Big),\label{eq:APP}
\end{align}
with
\begin{align}
a_1(\mu) &= C_2(\mu) + \frac{C_1(\mu)}{N_c}, \quad
a_2(\mu) = C_1(\mu) + C_2(\mu)[\frac{1}{N_c} + \chi^{C}e^{i\phi^{C}}],
\end{align}
and $N_c=3$. Here $P_1$ represents the pseudoscalar meson transited from the $D$ decays, and $P_2$ the emitted meson, in the $T$ and $C$ diagrams. $C_{1,2}(\mu)$ are the Wilson coefficients at the scale of $\mu = \sqrt{\Lambda m_D(1-r^2_2)}$ for $T$ and $C$ diagrams, and $ \mu = \sqrt{\Lambda m_D(1 - r^2_1)(1 - r^2_2)}$ for $E$ and $A$ diagrams, with $r_i=m_{P_i}/m_D$, to describe the $SU(3)$ breaking effect relating to the energy release of the final states. $\Lambda$ represents the momentum of the soft degree of freedom in the $D$ decays, fixed to be $\Lambda=0.5$GeV in this work. It has been shown that large nonfactorizable contributions exist in the $C$ diagram, resulting from the final-state interactions, which are parametrized as $\chi^{C} e^{i\phi^{C}}$. $f_{i}$ and $F_0$ are the decay constants and transition form factors, respectively, whose values are used as in \cite{FAT1,Fusheng:2011tw}. The $E$ and $A$ diagrams are dominated by the nonfactorizable contributions, parametrized as $\chi_{q,s}^{E,A}e^{i\phi_{q,s}^{E,A}}$, while the factorizable ones are neglected due to the helicity suppression. The subscripts $q$ and $s$ stand for the quark pairs produced from the vacuum as the $u$, $d$ quarks or the $s$ quark.
Due to the fact that the pion boson is a Nambu-Goldstone boson and quark-antiquark bound state simultaneously~\cite{lab26,lab9}, a strong phase factor $e^{iS_\pi}$ is introduced for each pion involved in the non-factorizable contributions of $E$ and $A$ amplitudes. In the end, all the non-factorizable parameters, $\chi^{C}$, $\phi^{C}$, $\chi_{q,s}^{E,A}$, $\phi_{q,s}^{E,A}$ and $S_{\pi}$ are universal parameters to
be fit from the data.

Similarly, the topological amplitudes of the $D\rightarrow PV$ modes can be parametrized as \cite{FAT2}
\begin{align}
 T_P[C_P] &= \frac{G_F}{\sqrt{2}}V_{CKM}a_1^P(\mu)[a_2^P(\mu)]f_Vm_VF_1^{D\rightarrow P}(m^2_V)2(\varepsilon_V\cdot p_D), \label{eq:TPCP}\\
  T_V[C_V] &= \frac{G_F}{\sqrt{2}}V_{CKM}a_1^V(\mu)[a_2^V(\mu)]f_P m_V A_0^{D\rightarrow V}(m^2_P)2(\varepsilon_V\cdot p_D), \label{eq:TVCV}\\
    E_{P,V} &= \frac{G_F}{\sqrt{2}}V_{CKM}C_2(\mu)\chi^E_{q,s}e^{i\phi^E_{q,s}}f_Dm_D\frac{f_Pf_V}{f_\pi f_\rho}(\varepsilon_V \cdot p_D),\label{eq:EPV}\\
 A_{P,V} &= \frac{G_F}{\sqrt{2}}V_{CKM}C_1(\mu)\chi^A_{q,s}e^{i\phi^A_{q,s}}f_Dm_D\frac{f_Pf_V}{f_\pi f_\rho}(\varepsilon_V \cdot p_D),\label{eq:APV}
\end{align}
where the subscript $P$ in $T_P$ and $C_P$ represents the topologies with a transited pseudoscalar meson and an emitted vector boson, while the subscript $V$ in $T_V$ and $C_V$ stands for the transited vector meson  and emitted  pseudoscalar meson diagrams.
The effective Wilson coefficients $a_1^{P(V)}$ and $a_2^{P(V)}$ are
\begin{equation}
\begin{split}
  a_1^{P(V)}(\mu) &= C_2(\mu) + \frac{C_1(\mu)}{N_C},    \quad
    a_2^{P(V)}(\mu) = C_1(\mu) + C_2(\mu)[\frac{1}{N_c} + \chi^C_{P(V)}e^{i\phi^C_{P(V)}}].
\end{split}
\end{equation}
The nonfactorizable parameters $\chi_{P,V}^C$ and $\phi_{P,V}^C$ are also free to be determined by the data. For the annihilation-type diagrams, the subscripts of $E_{P,V}$ and $A_{P,V}$ stand for the anti-quark from weak decays entering in the pseudoscalar meson or the vector meson. It is assumed that $E_{P}=E_{V}$ and $A_{P}=A_{V}$ in the FAT approach, due to the almost vanishing branching fraction of $D_{s}^{+}\to \pi^{+}\rho^{0}$~\cite{FAT2}, but $\chi_{q}^{E,A}\neq \chi_{s}^{E,A}$ and $\phi_{q}^{E,A}\neq \phi_{s}^{E,A}$ to describe large $SU(3)$ breaking effects.

In the end, the major nonperturbative and nonfactorizable contributions are involved in these universal parameters, and most $SU(3)$ breaking effects are considered in the FAT approach. Besides, the penguin contributions are not included in the CF and DCS decays, are smaller than the tree diagrams, and hence are neglected in this paper.  In the following discussions, the CKM matrix elements are specified out of each topological diagram to denote the CF and DCS amplitudes. So we will use the same symbols for the topological diagrams with and without CKM matrix elements, so there will be no ambiguity.

\section{Numerical results}\label{result}
\begin{table*}[t!]
\caption{Branching fractions and representations of topological amplitudes for the $D\to PP$ decays with $K_S^0$ or $K_L^0$ in the final states. Our results are given in the last column, compared to the experimental data~\cite{PDG}. }\label{tab:BrPP}
\begin{ruledtabular}
\footnotesize
\begin{tabular}{cccccc}
 Modes &   ~Representation~    &~
$\mathcal{B}_{\rm exp}(\%)$~& ~$\mathcal{B}_{\rm FAT}(\%)$~\\\hline
$D^0\to K_S^0\pi^0$ &  ${1\over2}V_{cd}^*V_{us}(C-E)-{1\over2}V_{cs}^*V_{ud}(C-E)$   & 1.20$\pm $0.04    & 1.31$\pm$0.06 \\
$D^0\to K_L^0\pi^0$ &$ {1\over2}V_{cd}^*V_{us}(C-E)+ {1\over2}V_{cs}^*V_{ud}(C-E)$  &1.00$\pm $0.07    & 1.05$\pm$0.04 \\
{$D^0\to K_S^0\eta$ } & \begin{tabular}{c} $ V_{cd}^*V_{us}[{1\over2}(C+E)\cos\phi_\eta -{1\over\sqrt2}E\sin\phi_\eta ]$ \\
 $~~~~~~~~~- V_{cs}^*V_{ud}[{1\over2}(C+E)\cos\phi_\eta -{1\over\sqrt2}E\sin\phi_\eta ]$ \end{tabular} & {0.485$\pm$0.030}   &{0.50$\pm$0.09}\\
$D^0\to K_L^0\eta$  & \begin{tabular}{c} $ V_{cd}^*V_{us}[{1\over2}(C+E)\cos\phi_\eta -{1\over\sqrt2}E\sin\phi_\eta ]$ \\
 $~~~~~~~~~+ V_{cs}^*V_{ud}[{1\over2}(C+E)\cos\phi_\eta -{1\over\sqrt2}E\sin\phi_\eta ]$ \end{tabular} &     &0.40$\pm$0.07  \\
$D^0\to K_S^0\eta'$ & \begin{tabular}{c} $ V_{cd}^*V_{us}[{1\over2}(C+E)\sin\phi_\eta +{1\over\sqrt2}E\cos\phi_\eta ]$ \\
 $~~~~~~~~~- V_{cs}^*V_{ud}[{1\over2}(C+E)\sin\phi_\eta +{1\over\sqrt2}E\cos\phi_\eta ]$ \end{tabular} & 0.95$\pm $0.05    &0.95$\pm$0.09  \\
$D^0\to K_L^0\eta'$ &\begin{tabular}{c} $ V_{cd}^*V_{us}[{1\over2}(C+E)\sin\phi_\eta +{1\over\sqrt2}E\cos\phi_\eta ]$ \\
 $~~~~~~~~~+ V_{cs}^*V_{ud}[{1\over2}(C+E)\sin\phi_\eta +{1\over\sqrt2}E\cos\phi_\eta ]$ \end{tabular} &     & 0.77$\pm$0.07\\
$D^+\to K_S^0\pi^+$ &$ {1\over\sqrt2}V_{cd}^*V_{us}(C+A)- {1\over\sqrt2}V_{cs}^*V_{ud}(T+C)$ & 1.53$\pm $0.06  &1.61$\pm$0.13  \\
$D^+\to K_L^0\pi^+$ &$ {1\over\sqrt2}V_{cd}^*V_{us}(C+A)+ {1\over\sqrt2}V_{cs}^*V_{ud}(T+C)$ & 1.46$\pm $0.05  &1.47$\pm$0.14  \\
$D_s^+\to K_S^0K^+$ & $ {1\over\sqrt2}V_{cd}^*V_{us}(T+C)- {1\over\sqrt2}V_{cs}^*V_{ud}(C+A)$& 1.50$\pm $0.05    &1.50$\pm$0.16  \\
$D_s^+\to K_L^0K^+$ & $ {1\over\sqrt2}V_{cd}^*V_{us}(T+C)+ {1\over\sqrt2}V_{cs}^*V_{ud}(C+A)$&     &1.46$\pm$0.16  \\
\end{tabular}
\end{ruledtabular}
\end{table*}

\begin{table*}[t!]
 \caption{Same as Table~\ref{tab:BrPP} but for the $D\to PV$ decays. }\label{tab:BrPV}
\begin{ruledtabular}
\footnotesize\begin{tabular}{ccccccc}
 Modes &   ~Representation~    &~
$\mathcal{B}_{\rm exp}(\%)$~& ~$\mathcal{B}_{\rm FAT}(\%)$~\\\hline
  $D^0\to K_S^0\rho^0$ & $ {1\over2}V_{cd}^*V_{us}(C_V-E_P)- {1\over2}V_{cs}^*V_{ud}(C_V-E_V)$& $0.64^{+0.07}_{-0.08}$  &$0.50\pm0.11$ \\
  $D^0\to K_L^0\rho^0$  &$ {1\over2}V_{cd}^*V_{us}(C_V-E_P)+ {1\over2}V_{cs}^*V_{ud}(C_V-E_V)$&  & $0.40\pm0.09$ \\
  $D^0\to K_S^0\omega$  &$ {1\over2}V_{cd}^*V_{us}(C_V+E_P)- {1\over2}V_{cs}^*V_{ud}(C_V+E_V)$&$1.11\pm0.06$ & $1.18\pm0.19 $ \\
  $D^0\to K_L^0\omega$  &$ {1\over2}V_{cd}^*V_{us}(C_V+E_P)+ {1\over2}V_{cs}^*V_{ud}(C_V+E_V)$&  & $0.95\pm0.15$\\
  $D^0\to K_S^0\phi$ & $ {1\over\sqrt2}V_{cd}^*V_{us}E_V- {1\over\sqrt2}V_{cs}^*V_{ud}E_P$& $0.424^{+0.033}_{-0.017}$  &$0.40\pm0.04$\\
  $D^0\to K_L^0\phi$ &$ {1\over\sqrt2}V_{cd}^*V_{us}E_V+ {1\over\sqrt2}V_{cs}^*V_{ud}E_P$&  & $0.33\pm0.03$ \\
  $D^+\to K_S^0\rho^+$  &$ {1\over\sqrt2}V_{cd}^*V_{us}(C_V+A_P)- {1\over\sqrt2}V_{cs}^*V_{ud}(T_P+C_V)$& $6.04^{+0.60}_{-0.34}$  &$4.99\pm0.50$ \\
  $D^+\to K_L^0\rho^+$  &$ {1\over\sqrt2}V_{cd}^*V_{us}(C_V+A_P)+ {1\over\sqrt2}V_{cs}^*V_{ud}(T_P+C_V)$&  &$5.37\pm0.50$ \\
  $D_s^+\to K_S^0K^{*+}$  &$ {1\over\sqrt2}V_{cd}^*V_{us}(T_P+C_V)- {1\over\sqrt2}V_{cs}^*V_{ud}(C_V+A_P)$& $2.7\pm0.6$  & $1.20\pm0.36$\\
  $D_s^+\to K_L^0K^{*+}$  &$ {1\over\sqrt2}V_{cd}^*V_{us}(T_P+C_V)+ {1\over\sqrt2}V_{cs}^*V_{ud}(C_V+A_P)$&   & $1.37\pm0.33$\\
\end{tabular}
\end{ruledtabular}
\end{table*}

In order to obtain the reasonable results of $K_{S}^{0}-K_{L}^{0}$ asymmetries $R(f)$, we do a global $\chi^{2}$ fit on the nonperturbative parameters in the FAT approach using the latest experimental data.
 The fittings are separately for the $D\to PP$ and $PV$ modes with 30 and 37 data, respectively.
We use the $\mathcal{B}(D\to K^0_{S}f)$ and $\mathcal{B}(D\to K^0_{L}f)$ instead of the $\mathcal{B}(D\to \overline K^0f)$ so as to include the interference effects between the CF and DCS decays.
The associated best-fit values and uncertainties are obtained as
\begin{equation}
  \begin{split}
\chi^{C} = -0.406\pm0.011, &~~~ \phi^C=0.636\pm0.011,\nonumber\\
\chi_q^E = 0.226\pm0.006, & ~~~\chi_s^E =0.138\pm0.005, \nonumber\\
\chi_q^A = 0.259\pm0.013, & ~~~\chi_s^A = 0.218\pm0.015, \nonumber\\
\phi^E_q = -4.44\pm0.02, & ~~~\phi_s^E = -4.81\pm0.04,\nonumber\\
\phi^A_q = -4.21\pm0.03, & ~~~\phi_s^A = -3.93\pm0.04,\nonumber\\
S_{\pi}=0.192\pm0.010, &
  \end{split}
\end{equation}
 in the $D\to PP$ modes, and
 \begin{equation}
\begin{split}
\chi_P^C = -0.443\pm0.007, & ~~~ \phi_P^C = 0.497\pm0.027, \nonumber\\
\chi_V^C = -0.694\pm0.024, &  ~~~\phi_V^C=0.828\pm0.065, \nonumber\\
\chi_q^E=0.194\pm0.013, &  ~~~\chi_s^E=0.283\pm0.011, \nonumber\\
\chi^A_q = 0.147\pm0.021, & ~~~\chi^A_s = 0.135\pm0.032,\nonumber\\
\phi_q^E = -1.40\pm0.07, & ~~~\phi_s^E = -3.09\pm0.13, \nonumber\\
\phi_q^A = -0.584\pm0.211, & ~~~\phi^A_s = -1.71\pm0.14,\nonumber\\
S_{\pi}=1.28\pm0.14, &
\end{split}
 \end{equation}
 in the $D\to PV$ modes.

The topological diagrammatic representations and our results of branching fractions in the $D\to K^0_Sf$ and $D\to K^0_Lf$ decays are presented in Tables~\ref{tab:BrPP} and \ref{tab:BrPV} for the $D\to PP$ and $PV$ decay modes, respectively. The predictions are given in the last columns,  compared to the experimental data~\cite{PDG}. The additional data and results in the global fitting are listed in Appendix. In order to obtain a reasonable error estimation, we consider the uncertainties of those universal parameters as well as the decay constants and form factors involved. The errors of decay constants of $\pi$, $K$, $D$ and $D_{s}$ are taken from PDG~\cite{PDG}, those of $\eta$ and $\eta'$ are from~\cite{Feldmann:1998vh}, and those of vector mesons are from~\cite{Straub:2015ica}. The form factors and their errors of $D\to P$ are taken from~\cite{Koponen:2012di}. The errors of all the other decay constants and form factors are taken as $10\%$ of the center value due to the theoretical uncertainties.
It can be found that our results are well  consistent with the data within the uncertainties. Besides, the predictions on the branching fractions of $D\to K_{L}^{0}f$ are to be tested by experiments.

From Tables. \ref{tab:BrPP} and \ref{tab:BrPV}, the branching fractions of the $D\to K_{S}^0f$ modes are obviously different from those of the $D\to K_{L}^0f$ modes, due to the effect of interference between the CF and DCS amplitudes. For example, $\mathcal{B}(D^0\to K_S^0f^0_{CP})$ are all larger than $\mathcal{B}(D^0\to K_L^0f^0_{CP})$. As shown in~\cite{Muller:2015lua}, $\mathcal{B}(D^0\to K_S^0\pi^0)>\mathcal{B}(D^0\to K_L^{0}\pi^0)$ holds with a significance of more than $4\sigma$. From (\ref{eq:ampKSKL}), $\mathcal{B}(D\to K_S^0f)+\mathcal{B}(D\to K_L^{0}f)=\mathcal{B}(D\to \overline K^0f)+\mathcal{B}(D\to K^{0}f)\approx\mathcal{B}(D\to \overline K^0f)$ as a good approximation with neglected branching fractions of DCS processes.

The difference between $\mathcal{B}(D\to K_S^0f)$ and $\mathcal{B}(D\to K_L^{0}f)$ is induced by the effect of interference between the CF and DCS amplitudes, defined by the $K^0_S-K^0_L$ asymmetries, $R(f)$.
With the fitting results,
the $K_{S}^{0}-K_{L}^{0}$ asymmetries in the $D^{+}$ and $D_{s}^{+}$ decays are predicted to be
\begin{equation}
\begin{split}
R(D^+\to K_{S,L}^{0}\pi^+) &=0.025\pm0.008, \\
R(D^+_s\to K_{S,L}^{0}K^+)  &=0.012\pm0.006 ,  \\
R(D^+\to K_{S,L}^{0}\rho^+) &= -0.037\pm0.011 ,   \\
R(D^+_s\to K_{S,L}^{0}K^{*+})  &= -0.070\pm0.032.
\end{split}
\end{equation}
Our result is consistent with the experimental data of
$R(D^+\to K_{S,L}^{0}\pi^+)_{\rm exp}=0.022\pm0.016\pm0.018$~\cite{lab10}.

For the $D^0$ decays, the amplitudes of $D^0 \rightarrow \overline{K}^0f^0_{CP}$ and $D^0 \rightarrow K^0f^0_{CP}$ are the same except for the CKM matrix elements.
For example,
\begin{equation}
\frac{\mathcal{A}(D^0\to K^0\pi^0)}{\mathcal{A}(D^0\to \overline{K}^0\pi^0)} =\frac{V^*_{cd}V_{us}}{V^*_{cs}V_{ud}} \frac{C_{K^0}+E_{K^0}}{C_{\overline{K}^0}+E_{\overline{K}^0}}=-\tan^{2}\theta_{C}.
\end{equation}
In the FAT approach, as showed in Eq.~\eqref{eq:TPP} and \eqref{eq:EPP},
$C_{K^0} = C_{\overline{K}^0}$, $E_{K^0} = E_{\overline{K}^0}$, due to $f_{K^{0}}=f_{\overline K^{0}}$. The above ratio is only related to the CKM matrix elements. This relation also holds for the $D^{0}\to K^{0}(\overline K^{0}) \eta^{(\prime)}$ decays. In the case of the $D\to PV$ modes, due to the assumption of  $E_{P}=E_{V}$ in the FAT approach as discussed in Sec. \ref{FAT} and shown in \eqref{eq:EPV}, the ratios between the DCS and CF amplitudes in the modes of $D^{0}$ decaying into  $\rho^{0}$, $\omega$ and $\phi$ also only depend on the CKM matrix elements.
Then the $K_{S}^{0}-K_{L}^{0}$ asymmetries in the $D^0$ decays are identical to each other, and according to Eq.~\eqref{eq:RD0}
\begin{equation}\label{p5}
 R(D^{0}\to K_{S,L}^{0}f_{CP}^{0})=2\tan^{2}\theta_{C}+y_{D}.
\end{equation}
With the current world averaging result of the $D^0-\overline D^0$ mixing parameter $y_{D}=(0.62\pm0.08)\%$ assuming no $CP$ violation~\cite{HFAG}, we have
\begin{equation}\label{eq:RD0num}
\begin{split}
 & R(D^0\to K^0_{S,L}\pi^0)=R(D^0\to K^0_{S,L}\eta)=R(D^0\to K^0_{S,L}\eta^{\prime})\\ &~~~~=R(D^0\to K^0_{S,L}\rho^0)
    =  R(D^0\to K^0_{S,L}\omega) =R(D^0\to K^0_{S,L}\phi) = 0.113\pm0.001,
\end{split}
\end{equation}
with the error from those of the CKM matrix elements and $y_{D}$. Our result is consistent with experimental result of $R(D^0\to K_{S,L}^{0}\pi^0)=0.108\pm0.025\pm0.024$~\cite{lab10} with large errors.
Without the effect of $D^{0}-\overline D^{0}$ mixing, $R(D^0\to K^0_{S,L}f^0_{CP})\approx0.107$ which is in agreement with predictions in other methods as seen in Table.\ref{tab:table1}. The improvement on the precision of measurements is called for to test the neutral $D$ mixing effect in the $K_{S}^{0}-K_{L}^{0}$ asymmetries.
In experiment, at the current stage with limited data to determine the effect of $D^{0}-\overline D^{0}$ mixing, it is suggested to measure all the above two-body decays of the $D^{0}$ and combine the results to decrease the errors.

It is found that the amplitudes of $\mathcal{A}(D^{0}\to K^{0} f_{CP}^{0})$ and $\mathcal{A}(D^{0}\to \overline K^{0} f_{CP}^{0})$ are reflected under the $U$-spin symmetry and the $K^0_S-K^0_L$ asymmetries of $D^0$ meson decays are less sensitive to the $SU(3)$ breaking, and thereby $R(D^0\to K^0_{S,L}\pi^0)=R(D^0\to K^0_{S,L}\eta)=R(D^0\to K^0_{S,L}\eta^{\prime})$~\cite{Rosner:2006bw}. Our results support this conclusion and extend it to $D^0\to PV$ decays. The results on the $PV$ modes depend on the assumption of $E_{P}=E_{V}$ in the FAT approach, which works well for the branching fractions at the current stage.

We have listed the results of the diagrammatic approach~\cite{ Bhattacharya:2009ps,Cheng:2010ry}, the QCD factorization approach~\cite{lab4}, the diagrammatic approach with global linear $SU(3)$ breaking analysis~\cite{Muller:2015lua}, the experimental data~\cite{lab10} and the FAT approach in Tables~\ref{tab:table1} for comparison.
Our prediction of  $R(D^0\to K_{S,L}^{0}\pi^0)$ is larger than the others by $y_{D}=(0.62\pm0.08)\%$ due to the $D^{0}-\overline D^{0}$ mixing effects involved.
The result of $R(D^+\to K_{S,L}^{0}\pi^+)$ in this work has the same sign with the experimental data, but opposite to the other theoretical predictions, because the FAT approach could contain significant flavor $SU(3)$ symmetry breaking effects compared with~\cite{Bhattacharya:2009ps,Cheng:2010ry,lab4}, and the latest experimental data of branching fractions have been considered. It is a similar case for the predictions of $R(D_s^+\to K_{S,L}^0K^+)$. In~\cite{Muller:2015lua}, since there are too many parameters to fit limited data, the uncertainties of predictions on the $K^0_S-K^0_L$ asymmetries are very large.

\begin{table*}[t!]
\caption{\label{tab:table1}Results on $K_S^0-K_L^0$ asymmetries in $D^0\to K_{S,L}^0\pi^0$, $D^+\to K_{S,L}^0\pi^+$ and $D_s^+\to K_{S,L}^0K^+$. Our results are compared to other approaches~\cite{ Bhattacharya:2009ps,Cheng:2010ry,lab4, Muller:2015lua} and the experimental data~\cite{lab10}.~~~~~~~~~~~~~~~~~~~~~}
\begin{ruledtabular}
\footnotesize\begin{tabular}{ccccccc}
  &$R$\cite{ Bhattacharya:2009ps}&$R$\cite{Cheng:2010ry} & $R$\cite{lab4} & $R$\cite{Muller:2015lua} & $R_{\text{exp}}$\cite{lab10} & $R(\text{FAT})$ \\
  \hline
   $D^0\to K_{S,L}^0\pi^0$ & $0.107$& $0.107$& $0.106$ & $0.09^{+0.04}_{-0.02}$ & $0.108\pm0.035$ & $0.113\pm0.001$ \\
  $D^+\to K_{S,L}^0\pi^+$ &$-0.005\pm0.013$& $-0.019\pm0.016$&$-0.010\pm0.026$ &  & $0.022\pm0.024$ & $0.025\pm0.008$ \\
  $D_s^+\to K_{S,L}^0K^+$ & $-0.002\pm0.009$ &$-0.008\pm0.007$& $-0.008\pm0.007$ & $0.11^{+0.04}_{-0.14}$ &  & $0.012\pm0.006$ \\
\end{tabular}
\end{ruledtabular}
\end{table*}
\section{Conclusions}\label{CON}
The effect of interference between the CF and DCS amplitudes results in the $K^0_S-K^0_L$ asymmetries in $D\to K^0_{S,L}f$ decays.
We present the formulas of the $K^0_S-K^0_L$ asymmetries, $R(f)$, and calculate them in the FAT approach in which significant nonperturbative effects and the $SU(3)$ asymmetry breaking effects are involved. The branching fractions of the decay modes with $K_{L}^{0}$ are predicted.
The results of $R(D^0\to K^0_{S,L}\pi^0)$ and $R(D^+\to K^0_{S,L}\pi^+)$ are in agreement with experimental data. We first predict the $K^0_S-K^0_L$ asymmetries in the decay modes with vector mesons in the final states.
Furthermore, we first consider the effect of $D^0-\overline D^0$ mixing in the study of $K^0_S-K^0_L$ asymmetries in neutral $D$-meson decays. It is found that $R(D^0\to K^{0}_{S,L}f^0_{CP})=2\tan^{2}\theta_{C} + y_{D}=0.113\pm0.001$, where $y_{D}$ is the $D^0-\overline D^0$ mixing parameter, with the value of $(0.62\pm0.08)\%$ and cannot be neglected. Our predictions will be tested by the future experiments with higher precision, like BESIII. Besides, we find all the $K^0_S-K^0_L$ asymmetries in the $D^{0}$ decays are identical to each other. Therefore it is suggested to measure all of them and combine the results to test the effect of $D^{0}-\overline D^{0}$ mixing.

\begin{acknowledgements}
This work is partially supported by National Natural Science Foundation of China under Grants No. 11347027 and No. 11505083 and the Fundamental Research Funds for the Central Universities under Grant No. lzujbky-2015-241.
\end{acknowledgements}

\begin{appendix}

\section{Branching fractions in global fitting}\label{ob}
We list the experimental data and our predictions of the channels we used to determine the universal parameters in the FAT approach in Tables~\ref{BrPP} and \ref{BrPV}. The global fitting of the $D\to PP$ and $D\to PV$ modes is separate. There are 30 observables to fix 11 free parameters in the $D\to PP$ modes and 37 observables to fix 13 free parameters in the $D\to PV$ modes.
The $\rho^0-\omega$ mixing,
\begin{equation}
|\rho^0\rangle=|\rho^0_I\rangle-\varepsilon|\omega_I\rangle,\qquad
|\omega\rangle=\varepsilon|\rho^0_I\rangle+|\omega_I\rangle,
\end{equation}
is considered in the $D\to PV$ modes to conform with the undated data of $\mathcal{B}(D^0\to \pi^0\omega)$ and $\mathcal{B}(D^+\to \pi^+\omega)$~\cite{b}, where $|\rho^0_I\rangle$ and $|\omega_I\rangle$ denote the isospin eigenstates of $\rho^0$ and $\omega$ and $\varepsilon$ is chosen to be $0.12$ as in \cite{FAT2}.

\begin{table*}[htp]
\caption{ The branching fractions we have used in the global fitting of $D\to PP$ modes, compared with our predictions.  All data in this table are obtained from PDG~\cite{PDG}.  The $D\to K^0_{S,L}f$ modes are not included but listed in Table~\ref{tab:BrPP}. \label{BrPP}}\label{BrPP}
\begin{ruledtabular}
\footnotesize\begin{tabular}{ccc|ccc}
Modes & $\mathcal{B}_{\text{exp}}$& $\mathcal{B}_{\text{FAT}}$ & Modes & $\mathcal{B}_{\text{exp}}$& $\mathcal{B}_{\text{FAT}}$\\
\hline
$D^0\to \pi^+K^-$ &(3.93$\pm $0.04)$\%$&($3.82\pm0.96$)\% & $D_s^+\to \pi^+\eta$ &(1.70$\pm $0.09)$\%$ & ($1.96\pm 0.44$)\%\\
$D_s^+\to \pi^+\eta^{\prime}$ &(3.94$\pm $0.25)$\%$&($4.67\pm0.62$)\%& $D^0\to \pi^+\pi^-$ &(1.421$\pm $0.025)\textperthousand&($1.418\pm0.093$)\textperthousand\\
$D^0\to K^+K^-$ &(4.01$\pm $0.07)\textperthousand&($3.92\pm 0.95$)\textperthousand& $D^0\to K^0_SK^0_S$ &(0.18$\pm $0.04)\textperthousand&(0.20$\pm 0.03$)\textperthousand\\
$D^0\to \pi^0\pi^0$ &(0.826$\pm $0.035)\textperthousand&($0.707\pm0.029$)\textperthousand& $D^0\to \pi^0\eta$ &(0.69$\pm $0.07)\textperthousand&(0.99$\pm0.08 $)\textperthousand\\
$D^0\to \pi^0\eta^{\prime}$ &(0.91$\pm $0.14)\textperthousand&(0.66$\pm 0.04$)\textperthousand& $D^0\to \eta\eta$ &(1.70$\pm $0.20)\textperthousand&(1.27$\pm 0.25$)\textperthousand\\
$D^0\to \eta\eta^{\prime}$ &(1.07$\pm $0.26)\textperthousand&(1.43$\pm 0.21$)\textperthousand& $D^+\to \pi^+\pi^0$ &(1.24$\pm $0.06)\textperthousand&(1.04$\pm 0.07$)\textperthousand\\
$D^+\to K^0_SK^+$ &(2.95$\pm $0.15)\textperthousand&(3.06$\pm1.18 $)\textperthousand& $D^+\to \pi^+\eta$ &(3.66$\pm $0.22)\textperthousand&(2.80$\pm 0.42$)\textperthousand\\
$D^+\to \pi^+\eta^{\prime}$ &(4.84$\pm $0.31)\textperthousand&(3.89$\pm 0.22$)\textperthousand& $D_s^+\to \pi^0K^+$ &(0.63$\pm $0.21)\textperthousand&(0.69$\pm 0.03$)\textperthousand\\
$D_s^+\to K^0_S\pi^+$ &(1.22$\pm $0.06)\textperthousand&(1.04$\pm 0.13$)\textperthousand& $D_s^+\to K^+\eta$ &(1.77$\pm $0.35)\textperthousand&(0.91$\pm 0.20$)\textperthousand\\
$D_s^+\to K^+\eta^{\prime}$ &(1.8$\pm $0.6)\textperthousand&(3.1$\pm 0.4$)\textperthousand& $D^0\to \pi^-K^+$ &(1.399$\pm $0.027)\textpertenthousand&(1.550$\pm 0.086$)\textpertenthousand\\
$D^+\to \pi^0K^+$ &(1.89$\pm $0.25)\textpertenthousand&(1.73$\pm 0.13$)\textpertenthousand& $D^+\to K^+\eta$ &(1.12$\pm$0.18)\textpertenthousand&(0.67$\pm 0.17$)\textpertenthousand\\
$D^+\to K^+\eta^{\prime}$ &(1.83$\pm$0.23)\textpertenthousand&(1.72$\pm 0.19$)\textpertenthousand&  &&\\
\end{tabular}
\end{ruledtabular}
\end{table*}
\begin{table*}[htp]
\caption{  Same as Table \ref{BrPP} but for the $D\to PV$ decays, in which $\mathcal{B}(D^0\to \pi^0\omega)$ and $\mathcal{B}(D^+\to \pi^+\omega)$ are taken from~\cite{b}, $\mathcal{B}(D^0\to \eta\omega)$ from~\cite{R. Kass}, and the others are obtained from PDG~\cite{PDG}.  \label{BrPV}}\label{BrPV}
\begin{ruledtabular}
\footnotesize\begin{tabular}{ccc|ccc}
Modes & $\mathcal{B}_{\text{exp}}$& $\mathcal{B}_{\text{FAT}}$ & Modes & $\mathcal{B}_{\text{exp}}$& $\mathcal{B}_{\text{FAT}}$\\
\hline
$D^0\to \pi^+K^{*-}$ &($5.43\pm0.44$)\%& (5.72$\pm$0.80)\%& $D^0\to \pi^0\overline K^{*0}$ &($3.75\pm0.29$)\% &(3.75$\pm$0.27)\% \\
$D^0\to K^-\rho^+$ &($11.1\pm0.9$)\%& (10.6$\pm$0.6)\%& $D^0\to \eta\overline K^{*0}$ &($0.96\pm0.30$)\% &(0.39$\pm$0.13)\% \\
$D^+\to \pi^+\overline K^{*0}$ &($1.57\pm0.13$)\%&(1.71$\pm$0.33)\% & $D_s^+\to \pi^+\rho^0$&($0.020\pm0.012$)\% &(0.002$\pm$0.001)\% \\
$D_s^+\to \pi^+\omega$ &($0.24\pm0.06$)\%&(0.17$\pm$0.05)\% & $D_s^+\to \pi^+\phi$&($4.5\pm0.4$)\% &(3.4$\pm$0.7)\% \\
$D_s^+\to K^+\overline K^{*0}$ &($3.92\pm0.14$)\%& (4.06$\pm$0.50)\%& $D_s^+\to \eta\rho^+$&($8.9\pm0.8$)\% & (9.1$\pm$1.6)\%\\
$D_s^+\to \eta^{\prime}\rho^+$ &($5.8\pm1.5$)\%&(1.4$\pm$0.4)\%& $D^0\to \pi^+\rho^-$&($5.09\pm0.34$)\textperthousand &(4.34$\pm0.59$)\textperthousand \\
$D^0\to \pi^0\rho^0$ &($3.82\pm0.29$)\textperthousand&(4.06$\pm$0.29)\textperthousand & $D^0\to \pi^0\omega$&($0.117\pm0.035$)\textperthousand & (0.130$\pm$0.031)\textperthousand\\
$D^0\to \pi^0\phi$ &($1.35\pm0.10$)\textperthousand& (1.09$\pm$0.08)\textperthousand& $D^0\to \pi^-\rho^+$&($10.0\pm0.6$)\textperthousand &(9.4$\pm$0.6)\textperthousand \\
$D^0\to K^+K^{*-}$ &($1.62\pm0.15$)\textperthousand&(1.96$\pm$0.31)\textperthousand & $D^0\to K^-K^{*+}$& ($4.50\pm0.30$)\textperthousand& (4.76$\pm$0.31)\textperthousand\\
$D^0\to \eta\omega$ &($2.21\pm0.23$)\textperthousand& (1.92$\pm$0.35)\textperthousand& $D^0\to \eta\phi$&($0.14\pm0.05$)\textperthousand & (0.20$\pm$0.06)\textperthousand\\
$D^+\to \pi^+\rho^0$ &($0.84\pm0.15$)\textperthousand& (0.54$\pm$0.06)\textperthousand& $D^+\to \pi^+\omega$&($0.279\pm0.059$)\textperthousand &(0.326$\pm$0.108)\textperthousand \\
$D^+\to \pi^+\phi$ &($5.66^{+0.19}_{-0.21}$)\textperthousand& (5.60$\pm$0.44)\textperthousand& $D^+\to K^+\overline K^{*0}$&($3.84^{+0.14}_{-0.23}$)\textperthousand &(3.42$\pm$0.68)\textperthousand \\
$D^+\to K^0_SK^{*+}$ &($17\pm8$)\textperthousand&(5$\pm$1)\textperthousand & $D_s^+\to \pi^+K^{*0}$&($2.13\pm0.36$)\textperthousand &(3.04$\pm$0.53)\textperthousand \\
$D_s^+\to K^+\rho^0$ &($2.5\pm0.4$)\textperthousand& (2.1$\pm$0.3)\textperthousand& $D_s^+\to K^+\phi$&($0.164\pm0.041$)\textperthousand & (0.142$\pm$0.052)\textperthousand\\
$D^0\to \pi^-K^{*+}$ &($3.45^{+1.80}_{-1.02}$)\textpertenthousand& (4.44$\pm$0.31)\textpertenthousand& $D^+\to \pi^+K^{*0}$&($3.9\pm0.6$)\textpertenthousand &(3.7$\pm$0.3)\textpertenthousand \\
$D^+\to K^+\rho^0$ &($2.1\pm0.5$)\textpertenthousand&(2.2$\pm$0.4)\textpertenthousand & $D_s^+\to K^+K^{*0}$& ($0.90\pm0.51$)\textpertenthousand&(0.23$\pm$0.03)\textpertenthousand\\
\end{tabular}
\end{ruledtabular}
\end{table*}

\end{appendix}

\newpage


\begin{thebibliography}{99}

\bibitem{lab01}M.~Artuso, B.~Meadows, and A.~A.~Petrov,
Annu.\ Rev.\ Nucl.\ Part.\ Sci.\  {\bf 58}, 249 (2008).

\bibitem{PDG}
C.~Patrignani {\it et al.} (Particle Data Group), Chin.\ Phys.\ C {\bf 40}, 100001 (2016).

\bibitem{HFAG}
  Y.~Amhis {\it et al.},
  arXiv:1612.07233
  and online update at 
  http://www.slac.stanford.edu/xorg/hfag.

\bibitem{Kagan:2009gb}
  A.~L.~Kagan and M.~D.~Sokoloff,
  Phys.\ Rev.\ D {\bf 80}, 076008 (2009).

\bibitem{FAT1}  H.~-n.~Li, C.~-D.~Lu, and F.~-S.~Yu,
  Phys. Rev. D {\bf 86}, 036012 (2012).

\bibitem{FAT2}  Q.~Qin, H.~n.~Li, C.~D.~Lu, and F.~-S.~Yu,
  Phys.\ Rev.\ D {\bf 89}, 054006 (2014).

\bibitem{Cheng:2010ry}
  H.~Y.~Cheng and C.~W.~Chiang,
 Phys.\ Rev.\ D {\bf 81}, 074021 (2010).

\bibitem{Cheng:2012xb}
  H.~Y.~Cheng and C.~W.~Chiang,
 Phys.\ Rev.\ D {\bf 86}, 014014 (2012).

\bibitem{Cheng:2016ejf}
  H.~Y.~Cheng, C.~W.~Chiang, and A.~L.~Kuo,
Phys.\ Rev.\ D {\bf 93}, 114010 (2016).

\bibitem{Muller:2015lua}
  S.~M\"uller, U.~Nierste, and S.~Schacht,
 Phys.\ Rev.\ D {\bf 92}, 014004 (2015).

\bibitem{Bhattacharya:2008ss}
  B.~Bhattacharya and J.~L.~Rosner,
 Phys.\ Rev.\ D {\bf 77}, 114020 (2008).

\bibitem{Bhattacharya:2008ke}
  B.~Bhattacharya and J.~L.~Rosner,
 Phys.\ Rev.\ D {\bf 79}, 034016 (2009);
 {\bf 81}, 099903(E) (2010).

\bibitem{Bhattacharya:2009ps}
  B.~Bhattacharya and J.~L.~Rosner,
 Phys.\ Rev.\ D {\bf 81}, 014026 (2010).

\bibitem{Donoghue:1985hh}
  J.~F.~Donoghue, E.~Golowich, B.~R.~Holstein, and J.~Trampetic,
 Phys.\ Rev.\ D {\bf 33}, 179 (1986).

\bibitem{Buccella:1996uy}
  F.~Buccella, M.~Lusignoli, and A.~Pugliese,
 Phys.\ Lett.\ B {\bf 379}, 249 (1996).

\bibitem{lab3}  Z.~z.~Xing,
  Phys.\ Rev.\ D {\bf 55}, 196 (1997).

\bibitem{Falk:2001hx}
  A.~F.~Falk, Y.~Grossman, Z.~Ligeti, and A.~A.~Petrov,
 Phys.\ Rev.\ D {\bf 65}, 054034 (2002).

\bibitem{Cheng:2010rv}
  H.~Y.~Cheng and C.~W.~Chiang,
 Phys.\ Rev.\ D {\bf 81}, 114020 (2010).

\bibitem{Rosner:1999xd}
  J.~L.~Rosner,
  Phys.\ Rev.\ D {\bf 60}, 114026 (1999).

\bibitem{Bigi}
  I.~I.~Y.~Bigi and H.~Yamamoto,
  Phys.\ Lett.\ B {\bf 349}, 363 (1995).


\bibitem{lab10}  Q.~He {\it et al.} (CLEO Collaboration),
  Phys.\ Rev.\ Lett.\  {\bf 100}, 091801 (2008).

\bibitem{RBES} W. Zheng, in {\it Pos CHARM}, 075 (2016).

\bibitem{Gao:2006nb}
  D.~N.~Gao,
  Phys.\ Lett.\ B {\bf 645}, 59 (2007).

\bibitem{lab4}  D.~N.~Gao,
  Phys.\ Rev.\ D {\bf 91},  014019 (2015).

\bibitem{lab5}  M.~Beneke, G.~Buchalla, M.~Neubert, and C.~T.~Sachrajda,
  Phys.\ Rev.\ Lett.\  {\bf 83}, 1914 (1999);

  Nucl.\ Phys.\ B {\bf 591}, 313 (2000).

\bibitem{lab6} Y.~-Y.~Keum, H.~-n.~Li, and A.~I.~Sanda,
  Phys.\ Lett.\ B {\bf 504}, 6 (2001);
  Phys.\ Rev.\ D {\bf 63}, 054008 (2001);
C.~-D.~Lu, K.~Ukai, and M.~-Z.~Yang,
  Phys.\ Rev.\ D {\bf 63}, 074009 (2001);
    C.~-D.~Lu and M.~-Z.~Yang,
  Eur.\ Phys.\ J.\ C {\bf 23}, 275 (2002).

\bibitem{lab02} C.~W.~Bauer, D.~Pirjol, and I.~W.~Stewart,
  Phys.\ Rev.\ Lett.\  {\bf 87}, 201806 (2001);
  Phys.\ Rev.\ D {\bf 65}, 054022 (2002).

\bibitem{Aaij:2016cfh}
  R.~Aaij {\it et al.} (LHCb Collaboration),
  Phys.\ Rev.\ Lett.\  {\bf 116}, 191601 (2016).

\bibitem{x1}   Z.~Z.~Xing,
  Phys.\ Lett.\ B {\bf 353}, 313 (1995).

\bibitem{Grossman:2006jg}
  Y.~Grossman, A.~L.~Kagan, and Y.~Nir,
  Phys.\ Rev.\ D {\bf 75}, 036008 (2007).

\bibitem{qaz} L.~-L.~Chau,
  Phys.\ Rept.\  {\bf 95}, 1 (1983).

\bibitem{lab22}G.~Buchalla, A.~J.~Buras, and M.~E.~Lautenbacher,
  Rev.\ Mod.\ Phys.\  {\bf 68}, 1125 (1996).

\bibitem{Fusheng:2011tw}
  F.~S.~Yu, X.~X.~Wang, and C.~D.~Lu,
  Phys.\ Rev.\ D {\bf 84}, 074019 (2011).

\bibitem{lab26}  H.~-n.~Li and S.~Mishima,
  Phys.\ Rev.\ D {\bf 83}, 034023 (2011).

\bibitem{lab9}  G.~P.~Lepage and S.~J.~Brodsky,
  Phys.\ Lett.\ B {\bf 87}, 359 (1979).

\bibitem{Feldmann:1998vh}
  T.~Feldmann, P.~Kroll, and B.~Stech,
  Phys.\ Rev.\ D {\bf 58}, 114006 (1998).

\bibitem{Straub:2015ica}
  A.~Bharucha, D.~M.~Straub, and R.~Zwicky,
  J. High Energy Phys. 08 (2016), 098.

\bibitem{Koponen:2012di}
  J.~Koponen {\it et al.} (HPQCD Collaboration),
  arXiv:1208.6242.

\bibitem{Rosner:2006bw}
  J.~L.~Rosner,
  Phys.\ Rev.\ D {\bf 74}, 057502 (2006).

\bibitem{b}  M.~Ablikim {\it et al.} (BESIII Collaboration),
  Phys.\ Rev.\ Lett.\  {\bf 116}, 082001 (2016).

\bibitem{R. Kass}  R. Kass, in {\it Proceedings of Europhysics Conference on High Energy Physics} (Krakow, Poland, 2009),
http://pos.sissa.it/cgi-bin/reader/conf.cgi?confid=84.

\end{thebibliography}
\end{document}